\documentclass[12pt,epsfig]{article}
\usepackage{amsmath}
\usepackage{amsfonts}
\usepackage{graphicx}
\usepackage[dvips]{epsfig}
\psfigdriver{dvips}
\usepackage{mathrsfs}

\begin{document}

%
\def\papertitlepage{\baselineskip 3.5ex \thispagestyle{empty}}
\def\preprinumber#1#2{\hfill \begin{minipage}{4.2cm}  #1
                 \par\noindent #2 \end{minipage}}
\renewcommand{\thefootnote}{\fnsymbol{footnote}}
\newcommand{\beq}{\begin{equation}}
\newcommand{\eeq}{\end{equation}}
\newcommand{\beqa}{\begin{eqnarray}}
\newcommand{\eeqa}{\end{eqnarray}}
\catcode`\@=11
\@addtoreset{equation}{section}
\def\theequation{\thesection.\arabic{equation}} 
\catcode`@=12
\relax
\newcommand{\Det}{\operatorname{Det}}
\newcommand{\Str}{\operatorname{Str}}
\newcommand{\tr}{\operatorname{tr}}

\newcommand{\ld}{\lambda}
\newcommand{\ks}{k\hskip-1.2ex /}
%
%
%
\papertitlepage
\setcounter{page}{0}
\preprinumber{KEK-TH-1074}{hep-th/0603189}
\baselineskip 0.8cm
\vspace*{2.0cm}
\begin{center}
{\large\bf Little IIB Matrix Model}
\end{center}
\vskip 4ex
\baselineskip 1.0cm
\begin{center}
           {Yoshihisa~Kitazawa${}^{1,2}$, Shun'ya Mizoguchi${}^{1,2}$ 
and Osamu Saito${}^{1,3}$} 
\\
\vskip 1em
       ${}^1${\it High Energy Accelerator Research Organization (KEK)} \\
       \vskip -2ex {\it Tsukuba, Ibaraki 305-0801, Japan} \\
\vskip 1em
       ${}^2${\it Department of Particle and Nuclear Physics} \\
       \vskip -2ex {\it The Graduate University for Advanced Studies} \\
       \vskip -2ex {\it Tsukuba, Ibaraki 305-0801, Japan}\\
\vskip 1em
      ${}^3$ {\it Institute for Cosmic Ray Research, University of Tokyo} \\
        \vskip -2ex {\it Kashiwa 277-8582, Japan}
\end{center}
\vskip 5ex
%
\baselineskip=3.5ex
\begin{center} {\bf Abstract} \end{center}
\vskip 2ex
We study the zero-dimensional reduced model of $D=6$ 
pure super Yang-Mills theory and argue that the large $N$ limit describes 
the $(2,0)$ Little String Theory.
The one-loop effective 
action shows that the force exerted between two diagonal 
blocks of matrices behaves as $1/r^4$, implying a six-dimensional spacetime.
We also observe that it is due to non-gravitational interactions. 
We construct wave functions and vertex operators 
which realize the $D=6$, $(2,0)$ tensor representation. 
We also comment on other ``little" analogues of the IIB matrix model and 
Matrix Theory with less supercharges. 

\vspace*{\fill}
\noindent
March 2006
\newpage
\renewcommand{\thefootnote}{\arabic{footnote}}
\setcounter{footnote}{0}
\setcounter{section}{0}
\baselineskip = 0.6cm
\pagestyle{plain}

\section{Introduction}
Realizing string theory as a matrix model is a powerful framework for understanding 
its nonperturbative aspects. The first successful example was the realization 
of $c<1$ string theories in terms of zero-dimensional bosonic matrix models 
in double scaling limits \cite{oldmatrixmodels}. 
More recently, among other approaches, 
the IIB matrix model \cite{IKKT} has been proposed as a nonperturbative formulation of 
type IIB string theory \cite{FKKT,AIKKT}. %
Basically, the IIB matrix model is defined as 
a zero-dimensional reduced model of $D=10$ super Yang-Mills theory, 
which may also be viewed as an 
effective theory of D-instantons.  See \cite{IIB} for a review and further references. 

The first link between the IIB matrix model 
and string theory is that the matrix model action can be regarded 
as that of a regularization of type IIB Green-Schwarz (GS) superstring 
in the Schild gauge.
Since the GS superstrings \cite{GS} can be defined classically in $D=6,4$ and $3$, and 
there also exist pure super Yang-Mills theories precisely in these dimensions,
one may ask what kind of theories are described if one considers 
zero-dimensional reduced models of less supersymmetric Yang-Mills theories.
Of course the GS superstrings in noncritical dimensions are known 
to suffer from Lorentz anomalies in the light-cone gauge quantization (See e.g.\cite{GSW}.).
It is not clear, however, what is an obvious inconsistency in the reduced models
because the identification can be made only by a classical argument.  
Therefore it is meaningful to ask what these theories are. 

In this paper we study 
the model obtained by 
dimensionally reducing $D=6$ (and also $D=4$) pure super Yang-Mills theory to zero 
dimensions\footnote{We call this matrix model with half as many supersymmetries 
``little IIB matrix model" 
for an obvious reason, anticipating possible connections to Little String Theory.}
with an emphasis on its string theory interpretation.
The Witten indices of such models were computed in \cite{MNS}. Using the topological 
formulation \cite{HiKa} some regularized correlation functions 
of certain operators were obtained \cite{KKN} and the grand canonical partition 
function was shown to be a tau function of the KP hierarchy. 
Also, these models  
were explored numerically in \cite{Nishimura}. %
We give evidence that the matrix model describes a six-dimensional $(2,0)$ 
supersymmetric theory without gravity and argue that the large-$N$ limit of the matrix 
model describes the $(2,0)$ Little String Theory (LST) \cite{LST}.  
We should note that Matrix Theory descriptions 
of little string theories in the infinite 
momentum frame have already been proposed 
and well-known \cite{ABKSS}; our proposal is another different, manifestly 
Lorentz covariant one in terms of a zero-dimensional reduced model.

In section 2 we first define our model, and compute the one-loop effective action.  
Unlike the maximally supersymmetric case, the force exerted between two diagonal 
blocks of matrices behaves as $1/r^4$, implying that it is a six-dimensional theory.
We also observe that it is due to non-gravitational interactions. 
We then construct vertex operators for this model, closely following \cite{Kitazawa_vertex,ITU}.
In section 3 we show that this ``little" matrix model realizes the $D=6$, $(2,0)$ chiral 
supersymmetry by constructing wave functions transforming as a $(2,0)$ tensor 
multiplet.
In section 4 we derive vertex operators for those particles in the $(2,0)$ tensor 
multiplet by expanding a supersymmetric Wilson loop operator.  
Finally in section 5 
we discuss relations between our model and Little String Theory. We also briefly  
comment on other ``little" analogues of the IIB matrix model and Matrix Theory.
Appendix summarizes the conventions of the $D=6$ symplectic Weyl spinors.

\section{The model}
Our starting point is the following matrix model action
\beqa
S&=&-\tr \left(\frac14{[}A_\mu,A_\nu{]}^2 +\frac12 \bar\psi^i \Gamma^\mu{[}A_\mu,\psi_i{]}
\right),
\label{S}
\eeqa
where $\mu,\nu=0,1,\ldots,5$. This action can be obtained by dimensionally reducing 
the $D=6$, $U(N)$  pure super Yang-Mills theory to zero dimensions, and as such
is the same form as the ordinary IIB matrix model action 
except the range of the space-time indices and the size of the gamma matrices. The 
matrices $\psi_i$ $(i=1,2)$ are symplectic Majorana-Weyl  spinors. 
The conventions used in this paper are summarized in Appendix.

The action is invariant under the $D=6$, $(2,0)$ supersymmetries 
\beqa
\bar\epsilon^i Q_i^{(1)}&=&i(\bar\epsilon^i \Gamma_\mu\psi_i)\frac\delta{\delta A_\mu}
-\frac i2{[}A_\mu, A_\nu{]}\Gamma^{\mu\nu}\epsilon_i
\frac\delta{\delta\psi_i},\nonumber\\
{\bar\xi}^i Q_i^{(2)}&=&\xi_i\frac\delta{\delta\psi_i}.
\label{Q2}
\eeqa
Again, all the spinor variables carry the symplectic Majorana index $i=1,2$, but 
except this, they are the same  as the transformations in the original IIB matrix 
model. 
If the contracted indices are suppressed (according to the 
NW-SE rule 
; for instance,
$
\bar\epsilon \Gamma_\mu\psi \equiv \bar\epsilon^i \Gamma_\mu\psi_i
$) 
then the transformations look completely identical.
 
The one-loop effective action of the IIB matrix model was computed in the original IKKT 
paper. We can use their result for our ``little" IIB matrix model with some trivial changes.
That is, we expand the matrices variables around the backgrounds 
$A_\mu=p_\mu$ and $\psi_i=0$ as
\beqa
A_\mu~=~p_\mu+a_\mu,~~~\psi_i~=~0+\chi_i
\eeqa
and integrate out  the fluctuations $a_\mu$ and $\chi_i$.
Then we obtain the one-loop effective action \cite{IKKT}
\beqa
W
&=&\frac12 \tr \log(P_\lambda^2{\delta_\mu}^\nu-2i {F_\mu}^\nu)
-\frac12\tr \log\left(
(P_\lambda^2-\frac i2 F_{\mu\nu}\Gamma^{\mu\nu})\frac{1+\Gamma_7}2
\right)
-\tr \log P_\lambda^2,
\eeqa
where $P_\mu$ denotes the adjoint representation matrix of $p_\mu$, and 
similarly $F_{\mu\nu}$ does that of $f_{\mu\nu}=i{[}p_\mu,p_\nu{]}$. $\frac{1+\Gamma_7}2$ is the projection operator onto 
the complex four-dimensional space of Weyl spinors of positive chirality in 
six dimensions. The leading term of the $1/P^2$ expansion of $W$ is
\beqa
W&=&\frac12 \tr \frac1 {P^2} F_{\mu\nu}
\frac1 {P^2} F^{\nu\mu}+O(P^{-4}).
\label{1-loop}
\eeqa

As in \cite{IKKT,Kitazawa_vertex} we assume that the background $p_\mu$ 
be in the block-diagonal form, and let $p_\mu^{(i)}$ 
and $f_{\mu\nu}^{(i)}$ be the $i$th block of 
$p_\mu$ and $f_{\mu\nu}$, respectively.
In this notation 
we can write 
the contribution $W^{(i,j)}$ from the $ij$-th block  as \cite{IKKT,Kitazawa_vertex}
\beqa
W^{(i,j)}&=&\frac 1{2 r^4}\Big(
\tr f_{\mu\nu}^{(i)} f^{(i)\nu\mu}\tr 1^{(j)}+(i\leftrightarrow j)
\nonumber\\
&&~~~~~
-2\tr f_{\mu\nu}^{(i)} \tr f^{(j)\nu\mu}
\Big)+\cdots,
\eeqa
where $r=|d_\mu^{(i)}-d_\mu^{(j)}|$ ($d_\mu^{(i)}$ is the ``center-of-mass"  of the $i$th block:
$p_\mu^{(i)}=d_\mu^{(i)}1^{(i)}+\mbox{traceless part} $.) 
is interpreted as the distance between the two diagonal blocks.

Unlike the maximally supersymmetric IIB matrix model, the expansion of $W$ 
starts with the quadratic term in $F_{\mu\nu}$, and 
consequently the force exerted from one block to another behaves like $1/r^4$, 
implying that the model describes an interaction in a six-dimensional spacetime. 
Moreover, we can see from the tensor structure that the first line  can be regarded as a 
scalar-scalar interaction, while the second line corresponds to a force due to 
exchanges of 2-form fields; there are no gravitational interactions to this 
order.  Thus we conclude that the lowest excitation of this matrix model is a $D=6$ tensor multiplet. 

\section{Wave functions}
In the previous section we have seen that the reduced model (\ref{S}) of $D=6$ pure 
super Yang-Mills theory is naturally regarded as describing some $D=6$, (2,0) 
supersymmetric  theory without gravity. In this and the next sections we will construct 
vertex operators for the particles in the (2,0) tensor multiplet. 
To this end we consider the following supersymmetric Wilson loop operator \cite{Hamada}
\beqa
w(\lambda,k)&=&\tr e^{\bar\lambda^i Q^{(1)}_i }e^{ik_\mu A^\mu}e^{-\bar\lambda^i Q^{(1)}_i }.
\label{w}
\eeqa
We assume that $k^2=0$. $\ld_i$ $(i=1,2)$ are symplectic Majorana Weyl spinors 
satisfying 
\beqa  
\Gamma_7\ld_i&=&+\ld_i.
\eeqa
After a straightforward calculation we end up with the commutation relations
\beqa
{[}
\bar\epsilon_1^i Q_i^{(1)},\bar\epsilon_2^j Q_j^{(1)}
{]}
&=&
{[}
(2\bar\epsilon_1^i\Gamma^\mu\epsilon_{2i})A_\mu,A_\nu
{]}\frac\delta{\delta A_\nu}+
{[}
(2\bar\epsilon_1^i\Gamma^\mu\epsilon_{2i})A_\mu,\psi_j
{]}\frac\delta{\delta \psi_j},
\nonumber\\
{[}
\bar\epsilon^i Q_i^{(1)},\bar\xi^j Q_j^{(2)}
{]}
&=&
-i\bar\epsilon^i\Gamma_\mu\xi_i
\frac\delta{\delta A_\mu}
\eeqa
up to the equations of motion.
The right hand side of the first line is a gauge transformation and hence vanishes 
on any gauge invariant operator. Using these relations we obtain
\beqa
e^{\bar\epsilon^i Q^{(1)}_i }w(\ld,k)e^{-\bar\epsilon^i Q^{(1)}_i }
&=&
w(\ld+\epsilon,k),\nonumber\\
e^{\bar\xi^i Q^{(2)}_i }w(\ld,k)e^{-\bar\xi^i Q^{(2)}_i }
&=&
e^{(\bar\xi^i\Gamma_\mu\ld_i)k^\mu} w(\ld,k),
\eeqa
or 
in their infinitesimal forms
\beqa
{[}\bar\epsilon^i Q^{(1)}_i ,w(\ld,k){]}
&=&
\epsilon_i\frac{\partial}{\partial \ld_i}w(\ld,k),\\
{[}\bar\xi^i Q^{(2)}_i ,w(\ld,k){]}
&=&
(\bar\xi^i\ks \ld_i) w(\ld,k).
\eeqa

The parameter $\ld_i$ may be thought of as an isolated eigenvalue of the 
matrix $\psi_i$ representing the whole effect of the background as a mean field \cite{ITU}
(See also \cite{ISTU}.);
$k_\mu$ is the Fourier transform of the similarly isolated eigenvalue of $A_\mu$.

We would like to have a vertex operator $V_f$ for a particle of 
wave function  
$f$ satisfy \cite{vertex,vertex2}
\beqa
\delta V_f &=& V_{\delta f}
\eeqa
under the supersymmetry. Therefore we first construct a representation of the 
$D=6$, $(2,0)$ superalgebra  
\beqa
\delta^{(1)}f(\ld,k)&=&{\epsilon}_i \frac\partial{\partial\ld_i} f(\ld,k)\nonumber\\
&=&{\bar\epsilon}^{i} \frac\partial{\partial\bar\ld^i} f(\ld,k),\\
\delta^{(2)}f(\ld,k)&=&({\bar\xi}^i \ks \ld_i)f(\ld,k),\nonumber\\
&=&-(\bar\ld^i \ks {\xi}_i)f(\ld,k)\label{del2}
\eeqa
in the space of polynomials of $\ld_i$ and find wave functions of the supermultiplet. 
Then if we expand the Wilson loop operator $w(\ld,k)$ (\ref{w}) in terms of those wave 
functions we can (in principle) 
automatically 
obtain desired vertex operators 
as their coefficients.

Let us start from the scalar
 wave function $\Phi=1$. Applying $\delta^{(2)}$ to it, we have 
\beqa
\delta^{(2)}\Phi&=&{\bar\xi}^i \ks \ld_i,
\eeqa
so we define the spinor wave function $\Psi_i$ as
\beqa
\Psi_i&=&\frac12\ks \ld_i.
\eeqa
Next we apply $\delta^{(2)}$ to $\Psi_i$ to find
\beqa
\delta^{(2)}\Psi_i&=&-(\bar\ld^j \ks {\xi}_j)\cdot \frac12\ks \ld_i
\nonumber\\
&=&-\frac1{32}k^\sigma k_\rho(\ld^j\Gamma_{\mu\nu\sigma}\ld_j)\Gamma^{\mu\nu\rho}
{\xi}_i
+\frac14(\bar\ld^j \ks \ld_i)\ks {\xi}_j
\eeqa
after some Fierz rearrangements summarized in Appendix.
Thus we have the wave functions of the 2-form field   
\beqa
B_{\mu\nu}&=&\frac12 b_{\mu\nu},
~~~
b_{\mu\nu}~=~k^\rho \bar\ld^i \Gamma_{\mu\nu\rho} \ld_i,
\eeqa
and another set of scalars 
\beqa
{\Phi^i}_j&=&\bar\ld^i \ks \ld_j. 
\eeqa
The field strength $H_{\mu\nu\rho}$ of $B_{\mu\nu}$ is manifestly self-dual:
\beqa
-\frac16\epsilon^{\mu\nu\rho\sigma\tau\ld}H_{\sigma\tau\ld}
&=&H^{\mu\nu\rho}
\eeqa 
since one can write it as
\beqa
H_{\mu\nu\rho}&=&3ik_{[\mu}B_{\nu\rho]}\nonumber\\
&=&\frac32 i\bar\ld^i \ks\Gamma_{\mu\nu\rho}\ks \ld_i.
\eeqa
$B_{\mu\nu}$ is further transformed as
\beqa
\delta^{(2)}B_{\mu\nu}&=&-\frac13(\bar\epsilon^i\Gamma_{\mu\nu}\ks\ld_j)(\bar\ld^j\ks\ld_i),
\eeqa
which leads us to the definition of the conjugate spinors
\beqa
\Psi^c_i&=&\frac13\ks \ld_j (\bar\ld^j\ks\ld_i).
\eeqa
Finally we choose 
\beqa
\Phi^c&=&\frac13(\bar\ld^i\ks\ld_j)(\bar\ld^j\ks\ld_i)
\eeqa
as the conjugate scalar wave function.

With these definitions one may check that these wave functions satisfy the following 
$D=6$, $(2,0)$ superalgebra 
\cite{algebra}
\beqa
\delta B_{\mu\nu}&=&-\bar\epsilon^I\gamma_{\mu\nu}\psi_I,\\
\delta \psi_I&=&+\frac i{48}H^+_{\mu\nu\rho}\Gamma^{\mu\nu\rho}\epsilon_I
+\frac i4\partial\hskip -1.2ex /{\phi_I}^J \epsilon_J,\\
\delta \phi^{IJ}&=& -4\bar\epsilon^{[I}\psi^{J]} -\Omega^{IJ}\bar\epsilon^K\psi_K
\eeqa
if they are identified with the fields $B_{\mu\nu}$, $\psi^I$ and $\phi^{IJ}$ as
\beqa
B_{\mu\nu}&=&\frac12 b_{\mu\nu},\nonumber\\
\psi_I~=~(\psi_i,\psi_{i'})
&=&(\Psi^c_i,\Psi_{i'}),\nonumber\\
{\phi^{12}}~=~\phi_{1'2'}&=&\Phi,\nonumber\\
{\phi^{i'}}_{j'}&=&{\Phi^{i'}}_{j'},\\
{\phi^{i}}_{j}&=&-{\Phi^{i}}_{j},\\
\phi^{1'2'}~=~\phi_{12}&=&-\Phi^c.
\eeqa
The identifications of the supersymmetry parameters are
\beqa
\epsilon^I=(\xi^i,-2\epsilon^{i'}),~~~
\bar\epsilon^I=(\bar\xi^i,-2\bar\epsilon^{i'}).
\label{eps_identification}
\eeqa
$I=(i,i')$, $J=(j,j'),\ldots$ ($i,j=1,2$ ; $i',j'=1,2$) are the $USp(4)$ indices. 
They are raised and lowered by multiplications of 
\beqa
\Omega^{IJ}=\left(\begin{array}{cc}
0&\varepsilon^{ij'}\\
\varepsilon^{i'j}&0
\end{array}\right)
,~~~
\Omega_{IJ}=\left(\begin{array}{cc}
0&\varepsilon_{ij'}\\
\varepsilon_{i'j}&0
\end{array}\right)
\label{Omega}
\eeqa
as
\beqa
\epsilon^I=\Omega^{IJ}\epsilon_J,~~~
\epsilon_I=\epsilon^J\Omega_{JI}.
\eeqa
The Majorana condition for the $USp(4)$ spinor $\ld_I$ is
\beqa
(\bar\ld^I)^T=C\ld^I.
\eeqa 
These rules are consistent with the definitions  (\ref{Omega}) and 
the identifications (\ref{eps_identification}). 
Due to (\ref{useful1})(\ref{useful2}), $\phi^{IJ}$ satisfy the 
constraints 
\beqa
\phi^{IJ}\Omega_{IJ}&=&0,~~~\phi^{IJ}~=~-\phi^{JI}
\eeqa
so there are only five independent scalars.

\section{Vertex operators}

In the previous section we obtained wave functions of the particles
in the $(2,0)$ tensor multiplet. In this section we construct vertex 
operators for these particles by expanding the supersymmetric Wilson 
loop in terms of wave functions. Vertex operators are given as coefficients
in this expansion:
\beqa
w(\lambda,k)&=&\Phi V_\Phi+\Phi_i V_{\Phi_i}
+B_{\mu\nu} V_{B_{\mu\nu}}+{\Phi^i}_j V_{{\Phi^i}_j}
+\Psi_i^c V_{\Psi_i^c}+\Phi^c V_{\Phi^c}.
\eeqa

We begin by rewriting the Wilson loop operator as follows:
\beqa
w(\lambda,k)&=&\tr e^{\bar\lambda^i Q^{(1)}_i }e^{ik_\mu A^\mu}e^{-\bar\lambda^i Q^{(1)}_i }
= \tr e^{\sum_{n=0}^{4} G_n},
\eeqa
where \hspace{25ex}
\scalebox{6}[1]{
\rotatebox{-90}{\{}} 
\\[-4ex]\hspace{43ex}$n$
\beqa
G_n=\frac{1}{n!}[\bar{\lambda} Q^{(1)},\cdot \cdot \cdot [\bar{\lambda} Q^{(1)},ik A]] .
\label{Gn}
\eeqa
In (\ref{Gn}) and below we suppress the indices contracted according to the NW-SE rule.     
Note that $G_n$ contains $n$ $\lambda$'s. The sum in eq. (4.1) terminates at fourth order because the on-shell 
$\lambda$'s have 4 independent components. Each $G_n$ can be evaluated as follows:
\beqa
  G_0&=&ikA, \nonumber\\
  G_1&=&-\left(\bar\lambda\ks\psi\right), \nonumber\\
  G_2&=&\frac{i}{4}[A_{\mu},A_{\nu}], \nonumber\\
  G_3&=&-\frac{1}{3!}b^{\mu\nu}[\bar\lambda \Gamma_{\mu}\psi,A_{\mu}], \nonumber\\
  G_4&=&\frac{1}{4}\left( \frac{1}{2}b^{\mu\nu}(\bar\lambda \Gamma_{\mu\rho\sigma}\lambda)
[[A^{\rho},A^{\sigma}],A_{\nu}]-ib^{\mu\nu}[\bar\lambda\Gamma_{\mu}\psi,\lambda\Gamma_{\nu}
\psi] \right),  \nonumber\\
  G_n&=&0 \hspace{0.5cm} (n \ge 5).
\eeqa
Expanding the exponential of Eq. (4.1) and collecting the terms with the same power of $\lambda$,
we can read off vertex operators. 

The leading order term, which has no $\lambda$, is
  $ \tr e^{ikA}.$
This should equal $\Phi$ vertex operator multiplied by $\Phi$ wave function, thus 
we obtain $\Phi$ vertex operator
\beqa
V_\Phi&=&\mbox{tr} ~e^{ikA}.
\eeqa

The first order term gives the $\Psi_i$ vertex operator $V_{\Psi_i}$ as 
\beqa
\tr e^{ikA} G_1=\tr e^{ikA} (\bar \psi \ks \lambda)
=V_{\Psi_i}\Psi_i.
\eeqa
Hence 
\beqa
V_{\Psi_i}&=&\mbox{tr}~ e^{ikA} 2\bar\psi^i.
\eeqa

The second order terms can be evaluated as follows:
\beqa
\Str e^{ikA}\left(
\frac1{2}G_1^2+G_2
\right)&=&\Str e^{ikA} \left(\left(-\frac{1}{32}k^{\rho}(\bar \psi \cdot\Gamma_{\rho\mu\nu}\psi)
         +\frac{i}{4}[A_{\mu},A_{\nu}]\right)b^{\mu\nu} 
          +\frac{1}{4}(\bar \psi^{i}\cdot\ks \psi_j)(\bar \lambda ^{j}\ks \lambda_i)\right) \nonumber\\
       &=&V_{B_{\mu\nu}}B_{\mu\nu}+V_{{\Phi^j}_i}{\Phi^j}_i,
\eeqa
where ``$\Str$" is the symmetrized trace (See \cite{ITU} for its definition and some properties.)
and $\cdot$ means that the operators are symmetrized. 
Thus we have the vertex operator for $B_{\mu\nu}$
\beqa
V_{B_{\mu\nu}}&=&\Str e^{ikA} \left(
-\frac1{16}k_\rho \bar\psi\cdot\Gamma^{\mu\nu\rho}\psi 
+\frac i2{[}A_\mu,~A_\nu {]}
\right)
\eeqa 
and for ${\Phi^j}_i$
\beqa
V_{{\Phi^j}_i}&=&\Str e^{ikA} \frac14 \bar\psi^i \cdot \ks \psi_j.
\eeqa

The $\Psi^c_i$ vertex operator can be obtained from the third order terms
\beqa
\lefteqn{ \Str e^{ikA}\left(\frac{1}{3!}G_1\cdot G_1 \cdot G_1+G_1\cdot G_2+G_3\right) }\nonumber\\
    &=&\Str e^{ikA}\left(\frac{1}{9}(\bar\psi^i\cdot \ks \psi_j)(\bar\psi^j\ks\lambda_l)
  (\bar\lambda^l \ks \lambda)-\frac{i}{6}\bar\psi^i\Gamma_{\mu\nu}[A^{\mu},A^{\nu}]\ks\lambda_j
  (\bar\lambda^j\ks\lambda_i)\right) 
  \nonumber\\
  &=& V_{\Psi^{c}_{i}} \Psi^{c}_{i}.
\eeqa
After some Fierz rearrangement and with a help of formulas for the 
symmetrized trace \cite{ITU} we find\beqa 
V_{\Psi_i^c}&=&\Str e^{ikA} \left(
\frac13(\bar\psi^i\cdot \ks\psi_j)\cdot\bar\psi^j
-\frac i2{[}A^\mu,~A^\nu{]}\cdot \bar\psi^i\Gamma_{\mu\nu} \right).
\eeqa

The computation becomes more complicated as the order of $\lambda$ becomes higher.
Although the vertex operator for $\Phi^c$ could also be read off from the 
fourth order terms, we use the following shortcut method: 
We first notice from (\ref{Q2})(\ref{del2}) that multiplying $\xi\ks\ld$ to $w(\lambda,k)$ is 
equivalent to replacing every $\psi_i$ with $\xi_i$. Since we have already computed what 
becomes of each wave function after the operation of $\delta^{(2)}$, we can use it to 
guess what the next-order vertex operator is, up to $\psi_i$-independent terms. 
The latter can also be determined by e.g. expanding the Wilson loop as above.
In this way we finally find the expression for the conjugate scalar vertex operator 
\beqa
V_{\Phi^c}&=&\Str e^{ikA}
\left(\frac1{48}(\bar\psi^i\cdot\ks\psi_j)(\bar\psi^j\cdot \ks\psi_i)
-\frac i{16}{[}A_\mu,~A_\nu{]}\cdot k^\rho\bar\psi\cdot\Gamma_{\mu\nu\rho}\psi
\right.\nonumber\\
&&\left.+\frac18{[}A_\mu,~A_\nu{]}\cdot{[}A^\nu,~A^\mu{]}
\right).
\eeqa

\section{Conclusions and discussion}

We have seen that the reduced model of $D=6$ super 
Yang-Mills theory appears to describe a theory with
(1) a six-dimensional spacetime,
(2) $D=6$, $(2,0)$ chiral supersymmetry, 
(3) a coupling to a self-dual 2-form field and
(4) no massless gravitons.
We have also consistently constructed wave functions and vertex 
operators transforming as a $(2,0)$ tensor multiplet, which we expect to 
describe emissions of the particles responsible for the above non-gravitational 
interactions. 
Technically the method we have described at the end of section 4 
can save much labor in computing vertex operators, and we expect that 
we can use it to derive the complete forms of vertex operators in the full 
IIB matrix model.

It seems that maximal supersymmetry is essential to include gravity 
in matrix models. On the other hand,  
the items (1)$\sim$(4) are the common features shared by the 
$(2,0)$ little string theory (LST) (See \cite{LSTreview1,LSTreview2} for reviews; also 
\cite{AFKS} for more recent discussions.). 
Basically a LST is defined as a decoupling limit of (5+1)-dimensional world 
volume theory on a stack of NS5-branes. 
Since the supersymmetry is (2,0)((1,1)) for type IIA(IIB) 
5-branes \cite{CHS}, the former is of our interest. 
It is believed to allow a 
holographic dual description in terms of strings on a linear dilaton background \cite{ABKS}.

Since matrix models in general are naturally expected 
to define (in the sense of t'Hooft) string theories in the large $N$ limit, it is tempting 
to conjecture that our model at large $N$ is a description of the $(2,0)$ LST. 
In this picture the number of 5-branes 
$k$ will correspond to the number of diagonal blocks, each size of which goes to 
infinity in the limit.
In support of this conjecture we note that, in addition to their  common
features they share as above, both our matrix model and (non-double-scaled \cite{GK}) 
LST have a single dimensionful parameter but no other dimensionless one.  
It is also consistent that we have 
successfully obtained a set of vertex operators for the $(2,0)$ tensor multiplet,
but the ones for the $D=6$ gravity multiplet cannot be constructed in this framework. 
Although all the evidence we have so far is still only a circumstantial one, 
the features are suggestive and worth to be explored.

In this paper we have studied the reduced model of $D=6$ super Yang-Mills, 
but the model reduced from $D=4$ is also interesting. In four 
dimensions there are also both the (classical) Green-Schwarz superstring and pure 
super Yang-Mills. Following the route of the original (or the little) IIB matrix model,
one can similarly obtain its action and
$D=4$, ${\cal N}=2$ supersymmetry.  We encounter the 
following two puzzles, however. 

One of them is the fact that the one-loop effective action similar 
to (\ref{1-loop}) starts with, again, the quadratic term  in $F_{\mu\nu}$
with the $r^{-4}$ 
factor, which would mean that the model lives in a six-dimensional spacetime. 
The other is that 
one cannot realize a  $D=4$, ${\cal N}=2$ supermultiplet in the polynomial space of 
$D=4$ Majorana spinors because it has only half as many degrees of freedom of
what are needed. This would mean that this matrix model simply describes
the initial $D=4$, ${\cal N}=1$ gauge theory, although there appears to be ${\cal N}=2$ 
supersymmetry. 
A better understanding of this 1/4-supercharge model  
is an interesting problem for future investigations.

In order to further examine the relation between the little IIB matrix models and LSTs, 
it will be useful to consider the problem in the dual linear-dilaton \cite{ABKS}
and/or the cigar $SL(2,R)/U(1)$ CFTs \cite{OV,M,ES,Murthy,HK,FNP,Kutasov,PS}. 
In particular, it was recently shown \cite{FNP2,AMT} that the world-volume theory of 
D-branes in a certain cigar CFT background is a lower-dimensional pure super 
Yang-Mills theory. Therefore, in view of the established role of $D=10$ super 
Yang-Mills theory in the critical superstring theories, it seems consistent that the 
little IIB matrix model, which is defined as a reduced model of a lower-dimensional 
theory, describes some non-critical string theory. 
It would be interesting if one can directly compare 
the correlation functions of the corresponding vertex operators in the little matrix models and 
their dual CFTs.

Finally, we will now briefly comment on ``little" analogues of Matrix Theory \cite{BFSS} 
(See \cite{Taylor}
for a review.) with less supercharges.
These models were studied in e.g. \cite{CH,HKK}.
Let us consider matrix quantum mechanics obtained by reducing, again, the $D=6$ and $4$ 
pure super Yang-Mills theories to one dimension, and compute one-loop effective 
actions around a two-particle background in the standard eikonal approximation 
\cite{BFSS,BB}. Namely, we set 
\beqa
B^1=\frac i2\left(
\begin{array}{cc}vt&0\\0&-vt
\end{array}
\right),~~~
B^2=\frac i2\left(
\begin{array}{cc}b&0\\0&-b
\end{array}
\right),~~~B^3=\cdots=B^{D-1}=0,
\eeqa
where $B^i$ $(i=1,\ldots,D-1)$ are the backgrounds of the matrix variables $X^i$ 
$(i=1,\ldots,D-1)$ of the ``little" Matrix Theory; they are the spacelike components
of the $D$-dimensional gauge field $A_\mu$. The computation of the one-loop effective 
action is completely the same as \cite{BB}, except for the numbers of various types of 
fields appearing in the action. 
The result is
\beqa
W&\equiv&
\prod\log\Det(\partial_\tau^2 +\mbox{mass}^2)\nonumber\\
&=&-\int_0^\infty\frac{ds}se^{-sb^2}\frac1{\sinh sv}
\left(c\cosh sv +b\cosh 2sv +\frac a2\right),
\eeqa
where  $a$, $b$ and $c$ are 
shown in Table 1.
(We also list the original Matrix Theory case ($D=10$) for comparison.)
Using these data, we find
$W=
\int_{-\infty}^\infty d\tau~
x\frac{v^2}{r^3}+O\Big(\frac{v^4}{r^{7}}\Big)$ 
%
%
with $x=\frac12$ $(D=6)$ and
$x=\frac34$ $(D=4)$.
\begin{figure}[b]
\begin{center}
Table 1.  The numbers of fields having different masses in the one-dimensional reduced 
models of  the $D=10$, $6$ and $4$ pure super Yang-Mills theories.
\\
\vskip 5mm
\begin{tabular}{|c||c|c|c|}
\hline
Number of fields & $D=10$ & $D=6$ & $D=4$ \\

\hline

$a$&$-6$&$-2$&$0$\\
$b$&$-1$&$-1$&$-1$\\
$c$&$+4$&$+2$&$+1$\\

\hline
\end{tabular}

\end{center}
\end{figure}

Note that the systematics of the expansions \cite{BBPT} in terms of $v$ and $b$ 
are valid in $D=6$ or $4$ without any change;
$v^2/r^3$ is the generic leading behavior of the potential and the above computations 
simply confirm that they do not vanish accidentally in the less supersymmetric cases.
This is a similar phenomenon to the divergence structure of super Yang-Mills theory 
\cite{FT}.
For the $D=6$ case, according to the conventional interpretation,
it suggests of some theory with  {\it seven} dimensional spacetime. One naturally 
thinks of it as describing ``little m theory" advocated in \cite{mandM} 
in the infinite-momentum frame, while 
a different interpretation of this model has been given in \cite{IPT}.
For the $D=4$ case, the long-range force again 
suggests of a seven-dimensional one rather than five. 
It will be interesting to compute brane charges \cite{BSS} 
for these little matrix models.

\section*{Acknowledgments}
We thank H.~Fuji, M.~Hatsuda, S.~Iso, H.~Kunitomo, 
J.~Nishimura, Y.~Okawa, H.~Suzuki, H.~Umetsu
and K.~Yoshida for useful discussions.
Y.~K. and S.~M. are supported in part by the Grant-in-Aid for Scientific Research
from the Ministry of Education, Science and Culture of Japan.

\renewcommand{\thesection}{A}
\setcounter{equation}{0}
\section*{Appendix}
The conventions of the gamma matrices are
\beqa
{\{}\Gamma^\mu,~\Gamma^\nu{\}}=2\eta^{\mu\nu},~~ ~
\eta^{\mu\nu}=\mbox{diag}(-1,+1,\ldots,+1).
\eeqa
\beqa
{\Gamma^0}^\dagger=-\Gamma^0,~~~
{\Gamma^i}^\dagger=+\Gamma^i~~(i=1,\ldots,5).
\eeqa
\beqa
\Gamma_7~=~\Gamma_0\Gamma_1\cdots\Gamma_5
~=~
-\Gamma^0\Gamma^1\cdots\Gamma^5.
\eeqa
The charge conjugation matrix $C$ satisfies
\beqa
C\Gamma^\mu C^{-1}~=~-{\Gamma^\mu}^T,~~~ C^T=+C.
\eeqa
Let $B=C\Gamma^0$, then 
\beqa
B\Gamma^\mu B^{-1}~=~+{\Gamma^\mu}^* ~~\mbox{(complex conjugate)}.
\eeqa
Let $\lambda$ be a complex Weyl spinor with chirality
\beqa
\Gamma_7\lambda=+\lambda.
\eeqa
\beqa
\bar\lambda=\lambda^\dagger\Gamma^0.
\eeqa

Any complex spinor $\lambda$ can be written as a sum of symplectic Majorana 
spinors 
\beqa
\ld=\ld_1+\ld_2, 
\eeqa
where 
\beqa
\ld_1&=&\frac12(\ld+B^{-1}\ld^*),\\
\ld_2&=&\frac12(\ld-B^{-1}\ld^*).
\eeqa
Then
\beqa
B\ld_1=\ld_2^*,~~~B\ld_2=-\ld_1^*.
\eeqa
Since $B$ commutes with $\Gamma_7$, this decomposition can be done in 
the subspace of spinors with definite chirality.
It is conventional to define
\beqa
\bar{\ld}^i=\ld_i^\dagger \Gamma^0~~~(i=1,2).
\eeqa
Note the positions of the indices. In this notation we have
\beqa
(\bar\ld^i)^T=C\ld^i,
\eeqa
where
\beqa
\ld^i=\varepsilon^{ij}\ld_j,~~~\ld_j=\ld^i \varepsilon_{ij},~~~\varepsilon^{12}=\varepsilon_{12}=+1.
\eeqa
The indices are raised (lower) by contracting $\varepsilon^{ij}$ ($\varepsilon_{ij}$) according to the 
NW-SE rule. Similarly decomposing another complex spinor $\epsilon$, we obtain the 
relation
\beqa
\frac12(\bar\epsilon\ks\ld-\bar\ld\ks\epsilon)&=&\bar\epsilon^i\ks\ld_i
\nonumber\\
&=&-\bar\ld^i\ks\epsilon_i.
\eeqa
The following relations are useful :
\beqa
\bar\ld^i\Gamma^\mu\ld_i&=&0,\label{useful1}\\
\bar\ld^i\Gamma^\mu\ld_j&=&+\bar\ld_j\Gamma^\mu\ld^i,\label{useful2}\\
\bar\ld^i\Gamma^{\mu\nu\rho}\ld_j&=&\frac12\delta^i_j\bar\ld^k\Gamma^{\mu\nu\rho}\ld_k.
\eeqa

For symplectic Majorana {\it Weyl} spinors $\ld_i$, $\psi_j$ with the same chirality, 
the Fierz rearrangement formula reads 
\beqa
\ld_j\bar\psi^i&=&-\frac14\Gamma^\mu (\bar\psi^i\Gamma_\mu \ld_j)
+\frac1{48}\Gamma^{\mu\nu\rho} (\bar\psi^i\Gamma_{\mu\nu\rho} \ld_j).
\eeqa

\end{document}